\documentclass[conference]{IEEEtran}
\usepackage[T1]{fontenc}
\usepackage{psfrag,amsmath,amssymb,color,cite, amsmath,graphicx}
\usepackage{stfloats}
\usepackage[caption=false,font=footnotesize]{subfig}
\usepackage{mathtools}

\DeclarePairedDelimiter\abs{\lvert}{\rvert}

\newcommand{\DeclareAutoPairedDelimiter}[3]{%
  \expandafter\DeclarePairedDelimiter\csname Auto\string#1\endcsname{#2}{#3}%
  \begingroup\edef\x{\endgroup
    \noexpand\DeclareRobustCommand{\noexpand#1}{%
      \expandafter\noexpand\csname Auto\string#1\endcsname*}}%
  \x}

\DeclareAutoPairedDelimiter\modulo{[}{]} 

\usepackage{algorithm}
\usepackage{mathrsfs}
\usepackage[algo2e,ruled,vlined]{algorithm2e} 
\usepackage{algpseudocode}
\usepackage{pifont}
\usepackage[colorinlistoftodos]{todonotes}

\newtheorem{lemma}{Lemma}

\newcommand{\beqn}{\begin{equation}}
\newcommand{\eeqn}{\end{equation}}
\newcommand{\beqa}{\begin{eqnarray}}
\newcommand{\eeqa}{\end{eqnarray}}
\newcommand{\beqas}{\begin{eqnarray*}}
\newcommand{\eeqas}{\end{eqnarray*}}

\presetkeys{todonotes}{color=green!30}{}
\usepackage{array}
\newcolumntype{P}[1]{>{\centering\arraybackslash}p{#1}}
\newcolumntype{M}[1]{>{\centering\arraybackslash}m{#1}}

\newcommand{\vy}{ {{\bf y}} }
\newcommand{\vx}{ {{\bf x}} }
\newcommand{\vw}{{{\bf w}} }
\newcommand{\vh}{{{\bf h}} }

\newcommand{\mh}{ {{\bf H}} }
\newcommand{\mx}{ {{\bf X}} }
\newcommand{\mg}{ {{\bf G}} }

\begin{document}

\title{Orthogonal Time Frequency Space (OTFS) Modulation Based Radar System}
\author{\IEEEauthorblockN{P. Raviteja\IEEEauthorrefmark{1}, Khoa T. Phan\IEEEauthorrefmark{2}, Yi Hong\IEEEauthorrefmark{1}, and Emanuele Viterbo\IEEEauthorrefmark{1}\\}
\IEEEauthorblockA{\IEEEauthorrefmark{1}ECSE Department, Monash University, Clayton, VIC 3800, Australia\\
\IEEEauthorrefmark{2}CSIT Department, La Trobe University, Bendigo, VIC 3550, Australia\\
Email: \{raviteja.patchava, yi.hong, emanuele.viterbo\}@monash.edu, k.phan@latrobe.edu.au}}

\maketitle

\begin{abstract}
Orthogonal time frequency space (OTFS) modulation was  proposed to tackle the destructive Doppler effects in wireless communications, with potential applications to many other areas. In this paper, we investigate its application to radar systems, and propose a novel efficient OTFS-based matched filter algorithm for target range and velocity estimation. The proposed algorithm not only exhibits the inherent advantages due to multi-carrier modulation of the existing orthogonal frequency division multiplexing (OFDM-) based radar algorithms but also provides additional benefits to  improve radar capability. Similar to OFDM, OTFS spreads the transmitted signal in the entire time--frequency resources to exploit the full diversity gains for radar processing. However, OTFS requires less cyclic prefix, and hence, shorter transmission duration than OFDM, allowing longer range radar and/or faster target tracking rate. Additionally, unlike OFDM, OTFS is inter-carrier interference-free, enabling larger Doppler frequency estimation.
We demonstrate the performance of the proposed algorithm using numerical results under different system settings.
\end{abstract}


\section{Introduction} 
The co-existence of radar sensing and wireless
communication (i.e., so-called RadCom) has found wide range of applications, both in modern civilian and commercial areas. For example, in emerging intelligent transportation applications, RadCom systems offer both communication links to other vehicles and active surrounding environment sensing functionalities, enabling cooperative interactions between all vehicles on the road  \cite{radar_sturm1}.  Other applications of RadCom systems are unsurprisingly found in aeronautical and military areas.  However, despite much research, there remain limitations of the current RadCom (and radar) systems to support emerging applications such as intelligent automotive systems with high mobility and dense traffic environment requiring very high data rate, ultra reliability, and ultra low latency communications. Under these conditions, it is challenging to develop suitable waveforms to simultaneously satisfy the requirements for both radar sensing and data communications.   

Orthogonal frequency division multiplexing (OFDM) modulation has been considered in existing literature for RadCom applications \cite{radar_sturm2,radar_gu,radar_turlapaty,radar_Wiesbeck,radar_tian,radar_Lellouch}.  Despite its many advantages such as simple random signal generation, full digital processing, and high processing gains in time and frequency, OFDM based waveforms exhibit drawbacks in radar sensing such as Doppler intolerance \cite{radar_Franken}. More critically, they also suffer from heavy degradation in data communications under high Doppler environments such as high-speed railway mobile communications.   

Recently, in \cite{Hadani,Hadani1}, the authors proposed orthogonal time frequency space (OTFS) modulation for communications over multi-path
delay--Doppler channels where each path exhibits a different delay and Doppler shift. The delay--Doppler domain provides as an alternative representation of a time-varying channel geometry due to moving objects (e.g. transmitters, receivers, or reflectors) in the scene. Leveraging on this representation, OTFS multiplexes each information symbol over a two dimensional (2D) orthogonal basis functions, specifically designed to combat the dynamics of time-varying multipath channels. Then the information symbols placed in the delay-Doppler coordinate system can be converted to the standard time-frequency domain used by traditional multi-carrier modulation schemes such as OFDM. 

OTFS was shown to provide significant error performance advantages over OFDM over delay--Doppler channels with a wide range of Doppler frequencies \cite{Hadani}--$\!\!$\cite{chocks}. With a suitable message passing based OTFS detection algorithm \cite{Ravi1}, the performance of OTFS is theoretically independent of Doppler frequencies unlike OFDM. These results inspire the potential applications of OTFS in many fields beyond wireless communications. In this paper, we investigate such an application of OTFS in RadCom and radar systems. While OTFS based waveforms are suitable for data communications, especially in high-mobility environments  such as self-driving cars, high-speed trains, drones, flying cars, and supersonic flights, it remains to be studied whether OTFS is well placed for radar processing (i.e., object angle, range, and velocity estimation), which is the focus of this paper. 

More specifically, this paper proposes a novel OTFS-based matched filter algorithm to  estimate the range and velocity of objects in radar systems. The proposed radar technique is motivated from our previous results on embedded channel estimation algorithms for OTFS \cite{Ravi3}, \cite{Ravi4}. Instead of embedding a known (or pilot) data symbol in an OTFS frame for transmit channel estimation, random equal power data symbols occupying a full OTFS frame are used for radar processing. By applying a simple matched filter-based algorithm on the received signal at the transmitter, we can efficiently estimate the number of potential targets as well as their corresponding ranges and velocities.  The proposed algorithm retains the inherent advantages of multi-carrier modulation  to exploit the full diversity gains for radar processing since OTFS spreads the transmitted signal in the entire time--frequency resources. Moreover, to transmit the same number of symbols, OTFS requires less cyclic prefix (CP), and hence, shorter transmission duration than OFDM, allowing longer range radar capability and/or faster target tracking rate. Another attractive property of OTFS-based radar is that it is inter-carrier interference (ICI-) free, enabling larger Doppler frequency estimation.  OFDM-based radar algorithms suffer larger ICI under higher Doppler frequencies due to the loss of orthogonality across sub-carriers, preventing the detection of larger Doppler frequencies (i.e., high-mobility targets). In general, while OFDM can exactly detect the Doppler frequencies only up to $10\%$ of the sub-carrier spacing $\Delta f$, OTFS can detect Doppler frequencies up to $\Delta f$. Therefore, we can conclude that OTFS-based radar is more robust to detect the long range and high velocity targets over conventional OFDM-based radar. 

The rest of the paper is organized as follows. Section II describes OTFS-based radar signal model, which lay the foundations for the development of OTFS-based matched filter radar algorithm in Section III. Numerical results are presented in Section IV followed by the conclusions in Section V. 

%%%%%%%%%%%%%%%%%%%%%%%%%%%%%%
\section{OTFS-based Radar} 

\begin{figure*}
\centering
\includegraphics[scale=0.5]{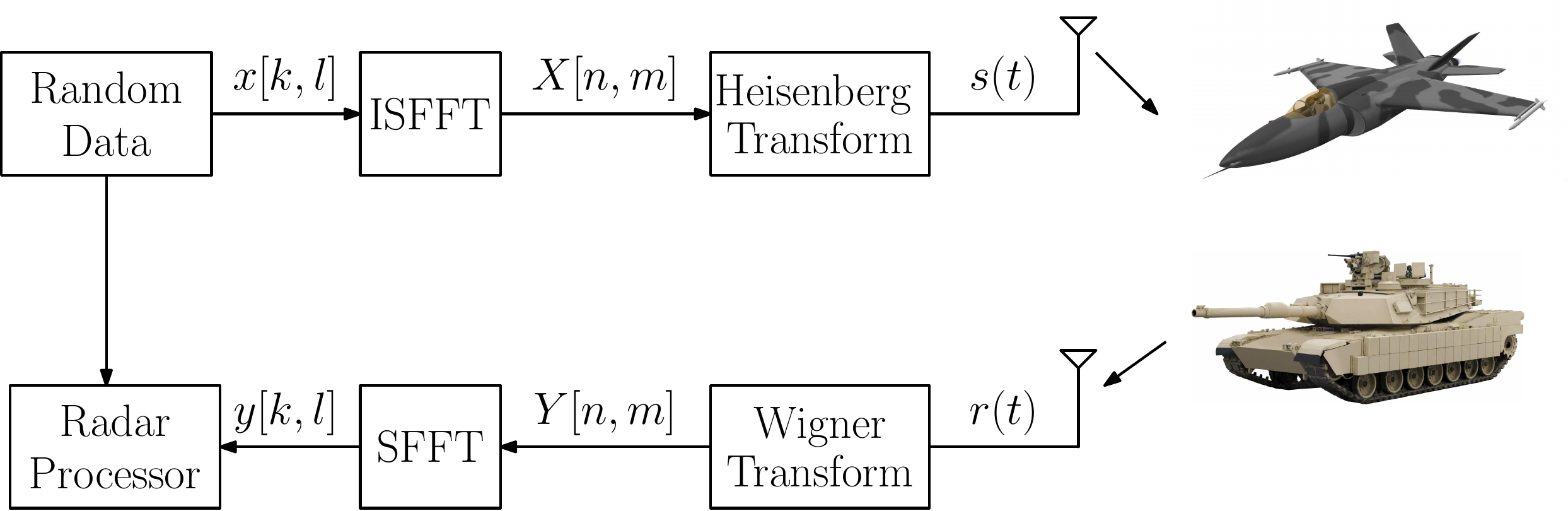}
\caption{OTFS-based radar system architecture}
\label{radar_bd}
\end{figure*}

In this section, we describe the signal model for OTFS-based radar following \cite{Hadani,Hadani1,Ravi1}.   
\subsection{Basic OTFS concepts/notations} 
-- The {\em time--frequency signal plane} is discretized to a $M$ by $N$ {\em grid} (for some integers $N, M >0$)  by sampling the time and frequency axes at intervals of $T$ (seconds) and $\Delta f=1/T$ (Hz), respectively, i.e., 
\beqn 
\Lambda = \bigl\{(nT,m\Delta f),\; n=0,\hdots,N-1, m=0,\hdots,M-1\bigr\} \nonumber 
\eeqn 

%-- A packet burst has duration $NT$ and bandwidth $M\Delta f$. 

-- The modulated {\em time--frequency samples} $X[n,m], n=0,\hdots,N-1, m=0,\hdots,M-1$ are transmitted over an OTFS frame with duration $T_f = NT$ and occupy a bandwidth $B = M\Delta f$. 

-- The {\em delay--Doppler plane} in the region $(0,T]\times (-\Delta f/2,\Delta f/2]$ is discretized to an $M$ by $N$ {\em grid} 
\beqn 
\Gamma = \Bigl\{\left(\frac{k}{NT},\frac{l}{M\Delta f}\right),\; k=0,\hdots,N-1, l=0,\hdots,M-1\Bigr\}, \nonumber 
\eeqn 
where $1/M\Delta f$ and $1/NT$  represent the quantization steps of the delay and Doppler frequency axes, respectively.

\subsection{OTFS-based radar signal model} 
The modulator first maps a set of $NM$ random symbols $\{x[k,l], k=0,\ldots,N-1, l=0,\ldots, M-1\}$ from a modulation alphabet $\mathbb{A} = \{ a_1, \cdots, a_{Q} \}$ (e.g. QAM symbols) of size $Q$, arranged on the delay--Doppler domain grid $\Gamma$, to samples $X[n,m]$  in the time--frequency domain  grid using the {\em inverse symplectic finite Fourier transform} (ISFFT). Next, the {\em Heisenberg transform} is applied to $X[n,m]$ using transmit pulse $g_{\rm tx}(t)$ to create the continuous-time {\em radar signal $s(t)$}, which is then transmitted over the multi-path delay--Doppler radar channel to be discussed in the following. 

Assume there are $P$ targets (or objects), each with range $R_i$ and relative velocity $V_i, i=1,\hdots,P$ relative to the transmitter. Note that $V_i$ can be positive or negative. Denote: 
$$
\frac{\tau_i}{2} = \frac{R_i}{c},\quad \frac{\nu_i}{2} = f_c\frac{V_i}{c}
$$
where $f_c$ is the carrier frequency, $c$ is the speed of light, $\tau_i$ is the round-trip delay between the transmitter and the $i$-th target, $\nu_i$ is the Doppler frequency of the $i$-th target. We can express the delay--Doppler radar channel $h(\tau,\nu)$ as: 
\beqn
h(\tau,\nu) = \sum_{i=1}^{P}h_i\delta(\tau - \tau_i)\delta(\nu - \nu_i), \label{radar}
\eeqn
where $h_i$ denotes the complex gain of the $i$-th target, and $\delta(\cdot)$ denotes the Dirac delta function. 
%We denote $l_{\tau_i}, k_{\nu_i}$ as the delay and Doppler {\em taps} for $i$-th path (relatively to %the delay--Doppler grid $\Gamma$): 
%\beqn
%\tau_i = \frac{l_{\tau_i}}{M\Delta f},\;\;\nu_i = \frac{k_{\nu_i}}{NT}
%\label{delaytap}
%\eeqn 
% Assume  $\tau_{\rm max}$, and $\nu_{\rm max}$ are the maximum delay and maximum Doppler shift among all channel paths.  Denote $l_{\tau}$ and $k_{\nu}$ the delay and Doppler taps corresponding to the largest delay $\tau_{\rm max}$ and Doppler $\nu_{\rm max}$.

As the grid dimension of the delay--Doppler plane is $M\times N$, the delay and Doppler ranges that can be detected in OTFS are given by $(0,\frac{1}{\Delta f}]$ and $(-\frac{1}{2T},\frac{1}{2T}]$, respectively. In this work, for simplicity, we assume the delay and Doppler of the targets are integer multiples of the delay resolution $\frac{1}{M\Delta f}$ and Doppler resolution $\frac{1}{NT}$, respectively.\footnote{We will consider the fractional multiples of the delay and Doppler in our future work.} Therefore, the delay--Doppler radar channel $h(\tau,\nu)$ in (\ref{radar}) will be expressed  as: 
\begin{align}
h(\tau,\nu) &\! = \!\sum_{k=0}^{N-1} \sum_{l=0}^{M-1} h[k,l]  \delta\!\left(\tau-\frac{l}{M\Delta f}\right)\! \delta\!\left(\nu-\frac{(k)_N}{NT}\right)  \label{radar1}
\end{align}
where 
\begin{align*}
 (k)_N & = \begin{cases}
 k & \mbox{if } k \leq N/2 \\
 k-N & \mbox{otherwise }
 \end{cases}
\end{align*}
%\todo{is it $\widetilde{k} = -k$ for $k \leq N/2$? -- it is $K-N$; $-1$ actual Doppler corresponds to $N-1$ $--$ circular shift}
and $h[k,l]$ denotes the complex gain of the target at the Doppler tap $k$ and delay tap $l$ corresponding to Doppler frequency $\frac{(k)_N}{NT}$ and delay $\frac{l}{M\Delta f}$. If there is not such target with Doppler tap $k$ and delay tap $l$ then  $h[k,l]=0$.
%Remind that the total number of targets are $P$ as in (\ref{radar}). } 

The time-domain signal $r(t)$ received at the transmitter after being reflected from the targets can be expressed as: 
\beqn
r(t) = \int \int h(\tau,\nu) s(t-\tau) e^{j2\pi\nu(t-\tau)}d\tau d\nu. \label{delay-doppler}
\eeqn
$r(t)$ is processed by the {\em Wigner transform} (implementing a receiver filter with an impulse response $g_{\rm rx}(t)$) followed by a sampler, to produce the received samples $Y[n,m]$ in the time--frequency domain. We then apply SFFT on $Y[n,m]$  to obtain the received symbols $y[k,l]$ in the delay--Doppler domain for radar processing.

We now look at the relations between received symbols $y[k,l]$ and transmitted symbols $x[k,l]$.
The exact relation between $y[k,l]$ and  $x[k,l]$, assuming rectangular pulses $g_{\text{tx}}(t)$ and $g_{\text{rx}}(t)$,  was derived in \cite{Ravi,Ravi2} as  
\begin{align}
y[k,l] = \sum_{k'=0}^{N-1} \sum_{l'=0}^{M-1} & h[k',l'] e^{j2\pi \left(\frac{[l-l']_M}{M}\right)  \frac{(k')_N}{N} }  \alpha_i[k,l]  \nonumber \\& x[[k - k']_N,  [l - l']_M] + w[k,l], \label{integer}
\end{align}
where 
\begin{align*}
\alpha_i[k,l] & = 
\begin{cases}
1 &  l' \leq l < M\\
e^{-j 2 \pi \frac{k}{N}} &  0\leq l<l'.
\end{cases}
\end{align*}
where  $w[k,l] \sim \mathcal{CN} (0, \sigma^2)$ is additive white noise with variance $\sigma^2$, $[\cdot]_N$ and $[\cdot]_M$ denote modulo $N$ and $M$ operations, respectively. We can see that the input--output relation in OTFS is similar to a 2D circular convolution except for an additional phase shift that depends on the location of symbols in the grid.
 
%%%%%%%%%%%%%%%%%%%%%%%%%%%%%%%%%%%%%%%%%%%%%%%%%%%%%%%%%%%%%%%%%%%%%%%%%%%
\section{OTFS-based Radar Matched Filter Algorithm}
%\subsection{An OTFS-based radar matched filter algorithm} 
In the OTFS-based radar system as in Figure 1, we   know the transmit symbols $x[k,l]$ and the received symbols $y[k,l]$. However, we do not know the delay--Doppler radar channel $h(\tau,\nu)$ in (\ref{radar}) or (\ref{radar1}). The purpose of radar processing is to estimate $h(\tau,\nu)$, which will provide us the information on the targets and corresponding ranges and velocities. 

We now propose a matched filter based method to estimate $h(\tau,\nu)$, which is equivalent to detect range and velocity of the targets, i.e., $h[k,l]$ for $0\leq k\leq N-1, 0\leq l\leq M-1$. %\textcolor{blue}{1. $h[k,l]$ is not defined previously. 2) How $h[k,l]$ relates to range and velocity of objects. }
%Here, we assume the maximum unambiguous delay and Doppler that radar can detect is equal to $1/\Delta f$ and $1/T$, respectively.
%with delay and Doppler resolutions $1/M\Delta f$ and $1/NT$

The input--output relation in (\ref{integer}) can be vectorized as
\begin{align}
\vy & = \mh \vx  + \vw,
\label{vec_y}
\end{align}
where $\mh \in \mathbb{C}^{MN\times MN}$ and  the $(k+Nl)$-th elements of $\vy,\vx,\vw\in \mathbb{C}^{MN\times 1}$ are equal to $y[k,l],x[k,l],$ and $w[k,l]$, respectively. %\textcolor{blue}{Should it be $v[k,l]$ instead of $w$?}. 
Then, we write (\ref{vec_y}) in an alternative form as
\begin{align}
\vy & = \widetilde{\mx} \vh + \vw,
\label{y_alternate}
\end{align}
\begin{figure*}[!t]
\normalsize
\begin{align}
\widetilde{\mx}[i,j] & = 
\begin{cases}
x[[k'-k'']_N,[l'-l'']_M] e^{-j2\pi\frac{k'}{N}} e^{j2\pi\frac{(k'')_N([l'-l'']_M)}{MN}} & \mbox{if }  l' < l''\\
x[[k'-k'']_N,[l'-l'']_M] e^{j2\pi\frac{(k'')_N([l'-l'']_M)}{MN}} & \mbox{otherwise}
\end{cases}
\label{X_eq}
\end{align}
\noindent\rule{16cm}{0.4pt}
%\vspace{-4mm}
\end{figure*}
where the $(k+Nl)$-th element of $\vh$ is $h[k,l]$ and the $(i,j)$-th element of $\widetilde{\mx} \in \mathbb{C}^{MN\times MN}$, $0\leq i=k'+Nl'\leq MN-1, 0\leq j=k''+Nl''\leq MN-1$ is given in (\ref{X_eq}).
%\todo{(\ref{X_eq}) in its current form does not make sense since the conditions should depends on $i, j$. Please check and consider revising. -- $i$ and $j$ only decides $k',l',k'',l''$ in (7)} 
Note that the first column of $\widetilde{\mx}$, i.e., $k''=l''=0$, is equal to $\vx$ and all the remaining columns of $\widetilde{\mx}$ are the circulant shifts of $\vx$ with a particular phase shift.

In order to obtain the estimate of $\vh$ from (\ref{y_alternate}), we propose the following matched filter (MF) detection: 
\begin{align}
\hat{\vh} = {\widetilde{\mx}}^{\rm H} \vy= {\widetilde{\mx}}^{\rm H} \widetilde{\mx}  \vh + {\widetilde{\mx}}^{\rm H} \vw  = \mg \vh + \widetilde{\vw} \label{mf_mg_eq}
\end{align}
where the gain matrix $\mg={\widetilde{\mx}}^{\rm H} \widetilde{\mx} \in \mathbb{C}^{MN\times MN}$ and $\widetilde{\vw} = {\widetilde{\mx}}^{\rm H} \vw$.
  
Assume the symbols $x[k,l], 0 \leq k \leq N-1, 0\leq l \leq M-1$ are drawn as independently and identically distributed (i.i.d.) quadrature phase shift keying (QPSK) symbols of power $P_s$. We can now examine the properties of the gain matrix $\mg$.

{1)} From the circulant property of $\widetilde{\mx}$, the $i$-th diagonal element of $\mg$, $\mg[i,i], 0 \leq i \leq MN-1$, reduces to
%all same and equal to $MN$.
\begin{align}
\mg[i,i] & = \sum_{k=0}^{N-1} \sum_{l=0}^{M-1} \abs{x[k,l]}^2 = MNP_s
\label{diag_elem}
\end{align}
%\todo{Should we use square brackets $\mg[i,i]$ instead of $\mg(i,i)$ for consistency?}
{2)} The off-diagonal elements of $\mg$, $\mg[i,j], 0 \leq i\neq j \leq MN-1$, can be written as
\begin{align}
\mg[i,j] = \sum_{k=0}^{N-1} & \sum_{l=0}^{M-1} x^*[[k-k']_N,[l-l']_M] \nonumber \\ 
& \hspace{0mm} x[[k-k'']_N,[l-l'']_M] \phi(i,j,k,l)
\label{off_diag_elem}
\end{align}
where $i=k'+Nl',j=k''+Nl''$ and
\begin{align}
\phi(i,j,k,l) & = \begin{cases}
\widetilde{\phi}(i,j,k,l) e^{-j2\pi \frac{k}{N}} & \mbox{if }  l' \leq l < l''\\
\widetilde{\phi}(i,j,k,l) e^{j2\pi \frac{k}{N}} & \mbox{if }  l'' \leq l < l'\\
\widetilde{\phi}(i,j,k,l) & \mbox{otherwise}
\end{cases}
\end{align}
where $\widetilde{\phi}(i,j,k,l) = e^{j2\pi \frac{(k'')_N([l-l'']_M)-(k')_N([l-l']_M)}{MN}}$.
Now, the mean and variance of $\mg[i,j]$ is given in the following lemma.

\begin{lemma}
The mean and variance of $\mg[i,j]$ are as follows
\begin{align}
\mathbb{E}[\mg[i,j]] & = 0 \\
\mbox{var}[\mg[i,j]] & = MNP_s^2
\end{align}
\label{lemma_mean}
\end{lemma}
{\em Proof:} See Appendix.
\hfill $\blacksquare$ \\
Hence, from (\ref{diag_elem}) and Lemma \ref{lemma_mean}, we can clearly see that as $MN$ increases, the normalized off-diagonal elements converge to zero in the mean-square sense, since 
\[\lim_{M,N\rightarrow \infty}\mbox{var}\left[\frac{1}{MNP_s}\mg[i,j]\right]=\frac{1}{MN}\rightarrow 0. \]
Therefore, for sufficiently large $MN$, (\ref{mf_mg_eq}) can be simplified to
\begin{align*}
\frac{1}{MNP_s} \hat{\vh} & \approx \vh + \frac{1}{MNP_s} \widetilde{\vw}.
\end{align*}
Here, the covariance matrix of $\frac{1}{MNP_s} \widetilde{\vw}$ is given by
\begin{align*}
%\mbox{covar}\left[\frac{1}{MNP_s} \widetilde{\vw}\right] =
\frac{1}{{(MNP_s)}^2} {\widetilde{\mx}}^{\rm H} \vw \vw^{\rm H} {\widetilde{\mx}}= \frac{\sigma^2}{{(MNP_s)}^2} \mg.
\end{align*}
%\todo{Not clear how this approximation is related to $\mg$. You might want to elaborate more on this. Note $\widetilde{\vw} = = {\widetilde{\mx}}^{\rm H} \vw$} 
Since $\mg$ is heavy diagonal, we can approximate $\frac{1}{MNP_s} \widetilde{\vw}$ as an i.i.d. Gaussian  noise vector having each entry distributed as $ \mathcal{CN} (0,\frac{\sigma^2}{MNP_s})$.

\begin{figure}
\centering
\subfloat[$M=4,N=4$]{
\includegraphics[height=1.8in,width=1.5in]{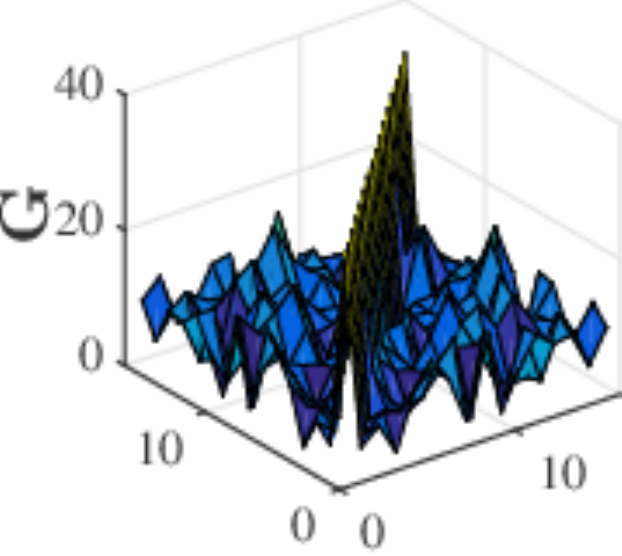}}
\subfloat[$M=32,N=32$]{
\includegraphics[height=2in,width=1.8in]{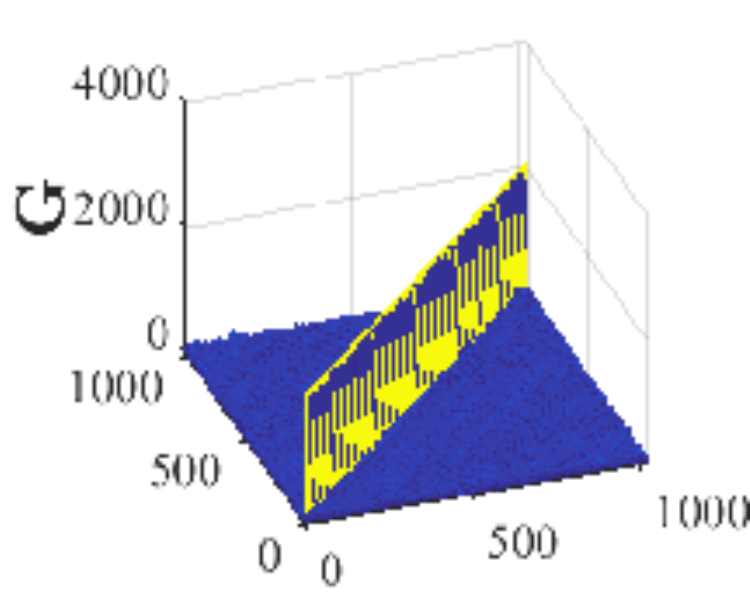}}
\caption{Magnitude plots of $\mg$ for $M=4,N=4$ and $M=32,N=32$}
\label{mag_fig}
\end{figure}

Figure \ref{mag_fig} shows the absolute value of $\mg$ for $M=4,N=4$ and $M=32,N=32$. From the figure, we can clearly see the diagonal dominance of the gain matrix $\mg$ for high values of $M$ and $N$.

Hence, through the MF detection, we can obtain multiple targets location that differ in at least one of the distance or relative velocity. The complexity of this detection is $\mathcal{O} ((MN)^2)$. 
\subsection{Advantages of OTFS-based radar over OFDM-based radar}
 OTFS-based radar processing has advantages over OFDM-based radar processing in two aspects namely time resources and high Doppler interference.
 
1) While OTFS-based radar system requires one CP to transmit all the $NM$ symbols, OFDM-based radar system \cite{radar_sturm1} requires $N$  CP's. Therefore, OTFS radar saves a total of $(N-1)L$ symbols transmission time, which is attractive particularly for the targets at the longer ranges. The reduction in transmission time also allows to detect the target for more number of times compared to OFDM which results in faster tracking rate.
%\todo{Elaborate more on why this helps longer range radar and/or faster tracking rate.}

%for the same range and Doppler detection of OFDM.\\
2) Under larger Doppler frequencies, OFDM system experience higher ICI due to the loss of orthogonality across subcarriers, limiting OFDM capability to detect the larger Doppler frequencies. In general, OFDM can exactly detect the Doppler frequencies only up to $10\%$ of the subcarrier spacing ($\Delta f$). However, OTFS can detect the Doppler frequencies up to $\Delta f$ without any interference.

%%%%%%%%%%%%%%%%%%%%%%%%%%%%%%%%%%%%%%%%%%%%%%%%%%%%
\section{Simulation Results and Discussion}
In this section, we demonstrate the superior performance of OTFS over OFDM through numerical results. For OFDM simulations, we consider the conventional FFT method proposed in \cite{radar_sturm1}.
The summary of the simulation parameters is given in Table \ref{tab1}. Here, we assume a single reflecting target with unit channel amplitude, i.e., $h_1=1$.

\begin{table}
\renewcommand{\arraystretch}{1.3}
\centering
\caption{Simulation Parameters}
  \label{tab1}
  \begin{tabular}[c]{ | c | M{3.7cm} | c | }
    \hline
    Symbol & Parameter & Value \\\hline
    $f_c$ & Carrier frequency & 24 GHz\\\hline
    $N_c$ & Number of subcarriers & 256 \\\hline
    $B$ & Total signal bandwidth & 10 MHz\\\hline
    $\Delta f$ & Subcarrier spacing & 39.063 KHz\\\hline
    $N_s$ & Number of evaluated symbols  & 64\\\hline
    $\Delta R$ & Range resolution & 15 m \\\hline
    $\Delta V$ & Velocity resolution & 3.8125 m/s \\\hline
    $R_{\rm max}$ & Unambiguous range & 3840 m \\\hline
    $V_{\rm max}$ & Unambiguous velocity & $\pm$ 122 m/s \\\hline
    SNR & Signal-to-noise ratio ($P_s/\sigma^2$) & 10 dB \\\hline
  \end{tabular}
\end{table}

\begin{figure}
\centering
\includegraphics[height=2.4in,width=3.2in,clip=true,trim=6mm 0mm 0mm 0mm]{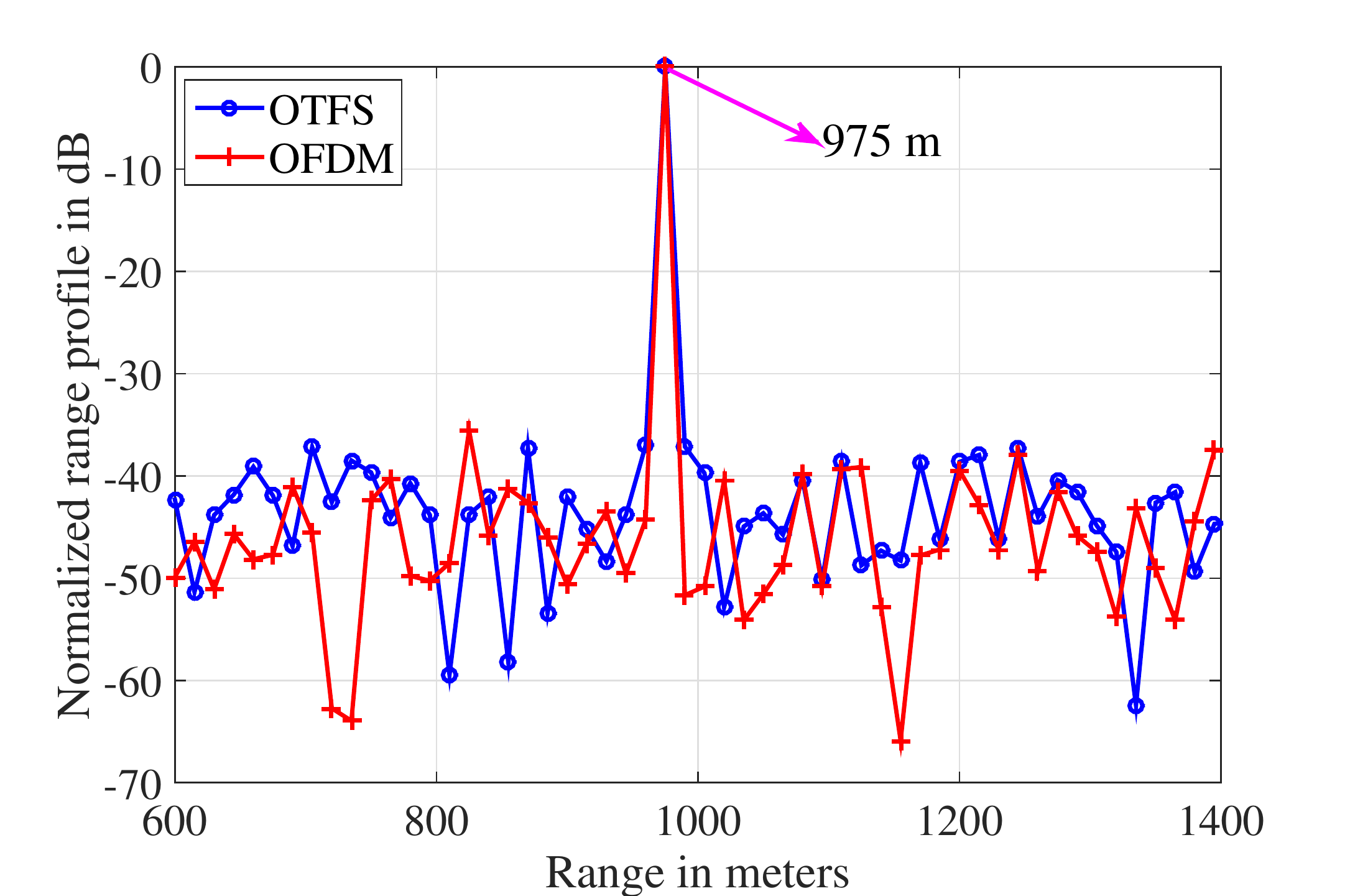}
\caption{OTFS vs OFDM radar range profile} 
\label{sim1}
\end{figure}

\begin{figure}
\centering
\includegraphics[height=2.4in,width=3.2in,clip=true,trim=6mm 0mm 0mm 0mm]{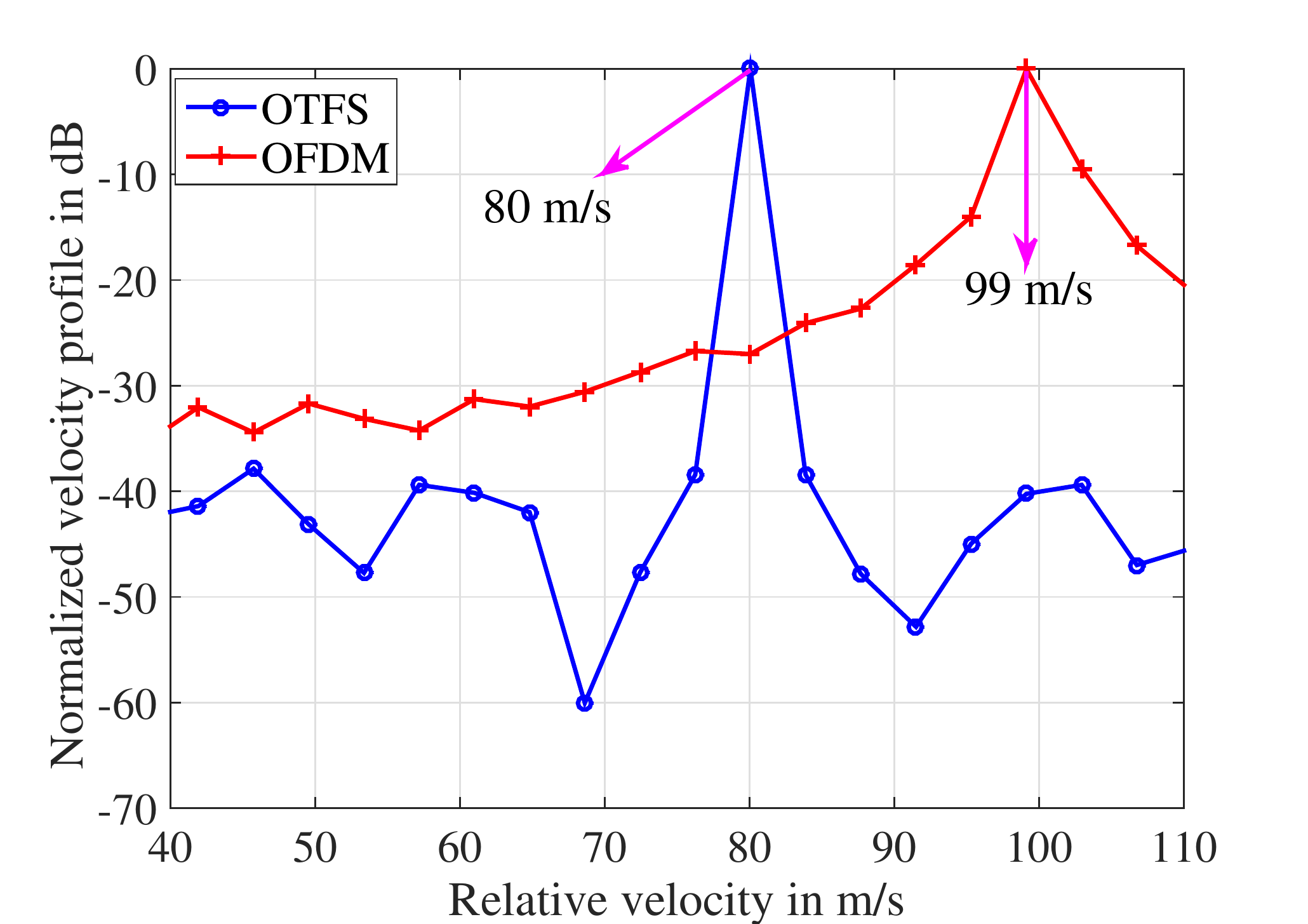}
\caption{OTFS vs OFDM radar Doppler profile}
\label{sim2}
\end{figure}

Figures \ref{sim1} and \ref{sim2} show the normalized range and velocity profile for the single reflecting object with $R=975$ m and $V = 80$ m/s. It can be observed that OFDM is able to detect the range of the target without any error, but experiences an error of $19$ m/s for the velocity. On the other hand, OTFS is able to detect both the range and velocity without any errors. Moreover, the peak-to-maximum sidelobe ratio (PSLR) is higher in OTFS compared to OFDM. The error in velocity and low PSLR of OFDM can be explained by the high ICI due to high Doppler ($65\%$ of $\Delta f$).  

\begin{figure}
\centering
\includegraphics[height=2.4in,width=3.2in,clip=true,trim=6mm 0mm 0mm 0mm]{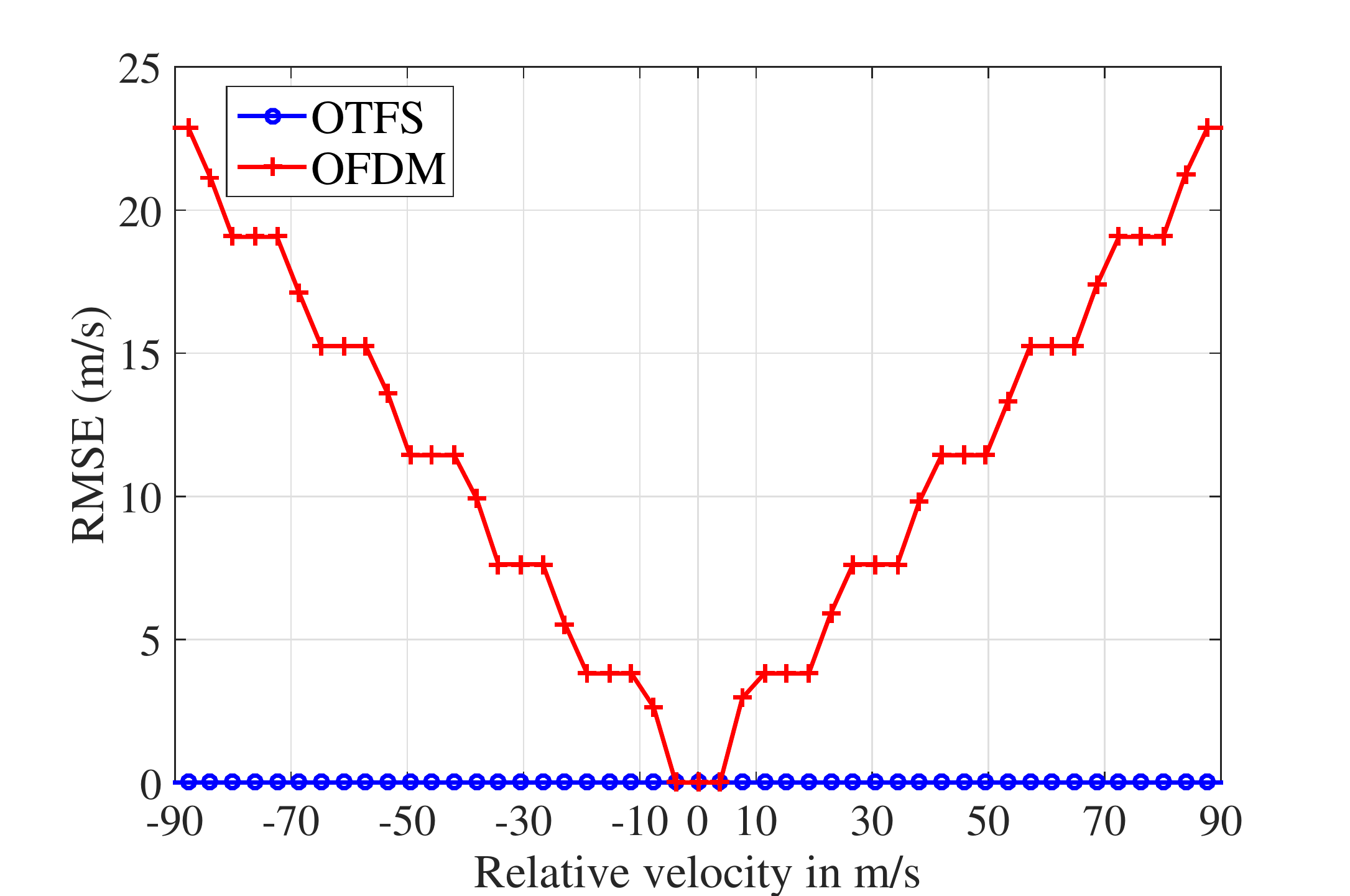}
\caption{RMSE vs relative velocity}
\label{sim3}
\end{figure} 

Figure \ref{sim3} displays the root mean squared errors (RMSE) of the velocity estimates for different target relative velocities with $R=975$ m. We assume the target velocities as the integer multiples of velocity resolution, $\Delta V$ and consider 100 Montecarlo simulations. We can see that while the RMSE of OFDM increases with relative velocity (> $25\%$ error at $\pm 90$ m/s), OTFS experience zero RMSE, which is useful to detect high Doppler targets.

\begin{figure}
\centering
\includegraphics[height=2.4in,width=3.2in,clip=true,trim=6mm 0mm 0mm 0mm]{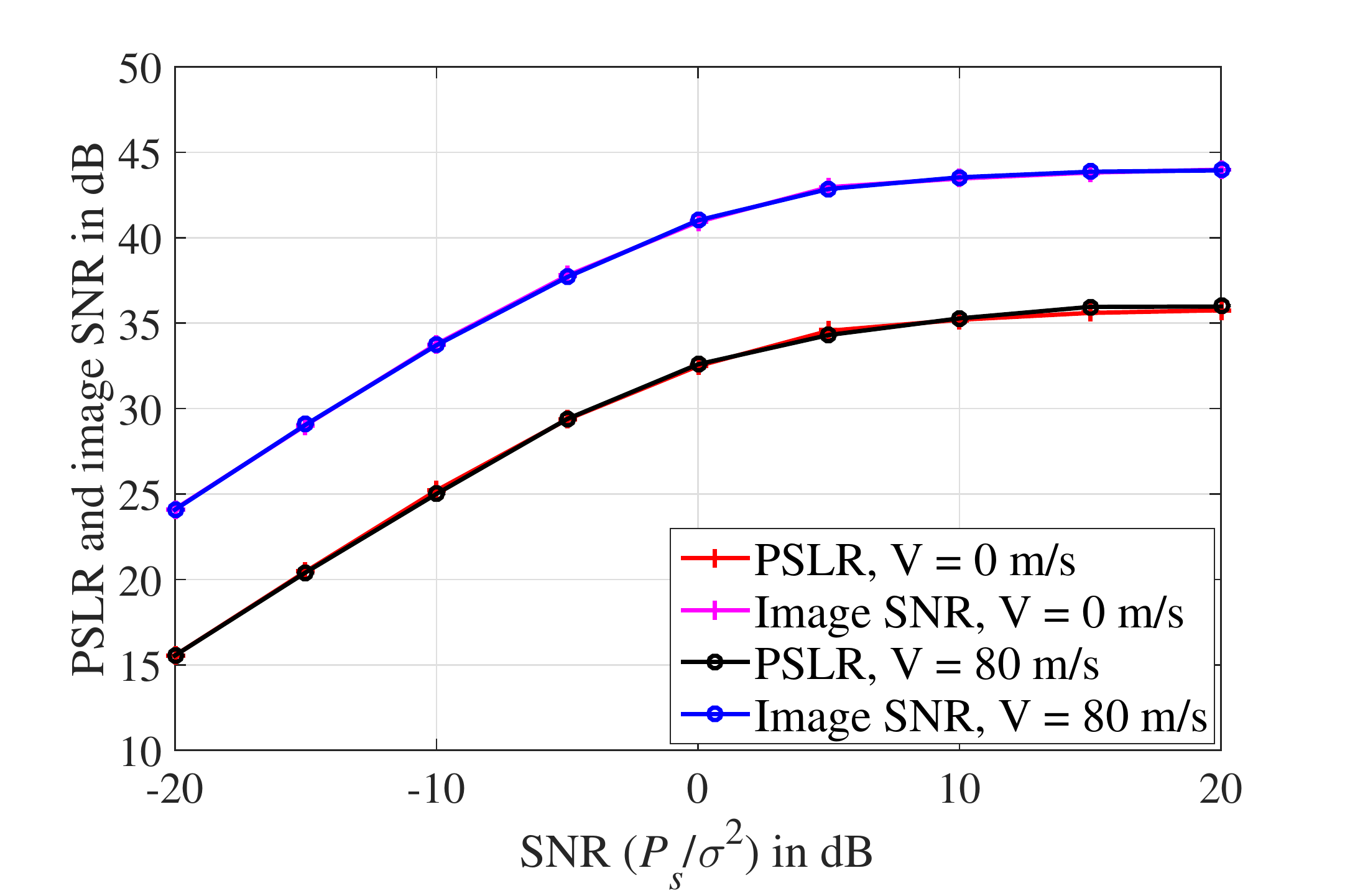}
\caption{PSLR and image SNR of OTFS radar}
\label{sim4}
\end{figure} 

Figure \ref{sim4} presents the average PSLR and image SNR of OTFS radar at different SNR $=P_s/\sigma^2$. The, image SNR represents the peak-to-average noise ratio with the radar processing gain and PSLR is the peak-to-maximum sidelobe ratio. It can be observed that the PSLR and image SNR of OTFS radar is independent of the relative velocity of target. Moreover, we see that at lower SNR's (typically below $0$ dB), the image SNR is close to the estimated value $\frac{\sigma^2}{MNP_s}$. Finally, at high SNR values, there is a saturation of image SNR to $\frac{1}{MN}$ which is due to the off-diagonal term of $\mg$. The saturation values of PSLR and image SNR's could be further improved by implementing more efficient detection algorithms, which we will consider in our future work.     

% \subsection{Future Work}

% -- OTFS pulse radar

% -- OTFS for fractional Doppler with different windows

% -- OTFS with better detection algorithms (sparse estimation)
%%%%%%%%%%%%%%%%%%%%%%%%%%%%%%%%%%%%%%%%%%%%%%%%%%%%
\section{Conclusion}
We have proposed a novel efficient orthogonal time frequency space  based matched filter algorithm for object range and velocity determination in radar systems. We show that OTFS-based radar processing not only exhibits the inherent advantages due to multi-carrier modulation  but also provides additional benefits for improved radar capability, such as longer range, faster tracking rate, as well as larger Doppler frequency estimation compared to the popular orthogonal frequency division multiplexing (OFDM) based radar. 
Our results demonstrate that OTFS-based radar with adequate detection algorithms is a promising robust technique to detect the long range and high velocity targets.

\section*{Acknowledgement}
This research work is supported by the Australian Research Council under Discovery Project ARC DP160101077. 

\appendix[Proof of Lemma \ref{lemma_mean}]
{\it Mean of $\mg[i,j]$} -- From (\ref{off_diag_elem}), the value of $\mathbb{E}[\mg[i,j]]$ can be simplified to zero as in from (\ref{mean_ind1}) to (\ref{mean_ind3}). Here, we invoke the i.i.d. property of $x[k,l]$ as $(k',l')\neq (k'',l'')$ for off-diagonal elements. 

\begin{figure*}
\begin{align}
\mathbb{E}[\mg[i,j]] & = \sum_{k=0}^{N-1} \sum_{l=0}^{M-1}   
\mathbb{E}\left[x^*[[k-k']_N,[l-l']_M] x[[k-k'']_N,[l-l'']_M]\right] \phi(i,j,k,l) \label{mean_ind1} \\
& = \sum_{k=0}^{N-1} \sum_{l=0}^{M-1} \mathbb{E}[x^*[[k-k']_N,[l-l']_M]] \mathbb{E}[x[[k-k'']_N,[l-l'']_M]] \phi(i,j,k,l) \label{mean_ind2}\\
& = 0 \label{mean_ind3}
\end{align}
\noindent\rule{16cm}{0.4pt}
\begin{align}
& \mbox{var}[\mg[i,j]] = \sum_{k_1=0}^{N-1} \sum_{l_1=0}^{M-1} \sum_{k_2=0}^{N-1} \sum_{l_2=0}^{M-1}  \mathbb{E}\Big[x^*[[k_1-k']_N,[l_1-l']_M] x[[k_1-k'']_N,[l_1-l'']_M] \nonumber \\
& \hspace{4cm} x[[k_2-k']_N,[l_2-l']_M] x^*[[k_2-k'']_N,[l_2-l'']_M] \Big] \phi(i,j,k_1,l_1) \phi^*(i,j,k_2,l_2) \label{var1} \\
& = \sum_{k_1=0}^{N-1} \sum_{l_1=0}^{M-1} \mathbb{E}\Big[\abs{x[[k_1-k']_N,[l_1-l']_M]}^2 \abs{x[[k_1-k'']_N,[l_1-l'']_M]}^2\Big] + \sum_{k_1=0}^{N-1} \sum_{l_1=0}^{M-1} \sum_{\substack{k_2=0\\k_2\neq k_1}}^{N-1} \sum_{\substack{l_2=0\\l_2\neq l_1}}^{M-1}  \mathbb{E}\Big[x^*[[k_1-k']_N,[l_1-l']_M] \nonumber \\
& \hspace{1cm} x^*[[k_2-k'']_N,[l_2-l'']_M] \Big] \mathbb{E}\Big[x[[k_1-k'']_N,[l_1-l'']_M] x[[k_2-k']_N,[l_2-l']_M] \Big] \phi(i,j,k_1,l_1) \phi^*(i,j,k_2,l_2) \label{var2} \\
& = MNP_s^2 + \sum_{k_1=0}^{N-1} \sum_{l_1=0}^{M-1} \sum_{\substack{k_2=0\\k_2\neq k_1}}^{N-1} \sum_{\substack{l_2=0\\l_2\neq l_1}}^{M-1} \mathcal{F}_{i,j}(k_1,l_1,k_2,l_2) \phi(i,j,k_1,l_1) \phi^*(i,j,k_2,l_2) = MNP^2 + \mathcal{F}_{i,j} \label{var3}
\end{align}
\noindent\rule{16cm}{0.4pt}
\begin{align}
\mathcal{F}_{i,j}(k_1,l_1,k_2,l_2) & =  \mathbb{E}\Big[{x^*}^2[[k_1-k']_N,[l_1-l']_M] \Big] \mathbb{E}\Big[x^2[[k_1-k'']_N,[l_1-l'']_M]\Big] \label{var4} \\
& = 0 \label{var5}
\end{align}
\noindent\rule{16cm}{0.4pt}
\end{figure*}

{\it Variance of $\mg[i,j]$} -- From (\ref{off_diag_elem}) and (\ref{mean_ind3}), the value of $\mbox{var}[\mg[i,j]]$ can be written as (\ref{var1}). The value of (\ref{var1}) can be split into sum of $MNP_s^2$ and $\mathcal{F}_{i,j}$ as in (\ref{var2}) and (\ref{var3}). The assumption $(k',l')\neq (k'',l'')$ is used in (\ref{var2}) and the i.i.d. property of $x[k,l]$ is used in (\ref{var3}).

Now, the value of $\mathcal{F}_{i,j}(k_1,l_1,k_2,l_2)$ may be non-zero only if the following four conditions are satisfied:
\begin{align*}
[k_1-k']_N = [k_2-k'']_N, [l_1-l']_M = [l_2-l'']_M\\
[k_1-k'']_N = [k_2-k']_N, [l_1-l'']_M = [l_2-l']_M
\end{align*}
In these cases, the value of $\mathcal{F}_{i,j}(k_1,l_1,k_2,l_2)$ can be simplified to zero as shown in (\ref{var4})--(\ref{var5}), which concludes the proof.  

% For simplicity of the proof, here we assume $k'>k''$ and $l'>l''$. The variance for the other cases can be straightforwardly extended.     
% After some algebraic manipulations, the above four conditions can be simplified to 
% \begin{align}
% k'=k''+\frac{N}{2}, l'=l''+\frac{M}{2}\\
% k_1=\modulo{k_2+\frac{N}{2}}_N, l_1=\modulo{l_2+\frac{M}{2}}_M
% \end{align}
% with either $l_1<l',l''$; $l''\leq l_2 < l'$ or $l''\leq l_1<l'$; $l_2 < l',l''$. 


\begin{thebibliography}{1}

\bibitem{radar_sturm1}
C. Sturm and W. Wiesbeck, ``Waveform design and signal processing aspects for fusion of wireless communications and radar sensing,'' in {\em Proc. IEEE.,} vol. 99, no. 7, pp. 1236-1259, Jul. 2011.

\bibitem{radar_sturm2}
C. Sturm, M. Braun, T. Zwick, and W. Wiesbeck, ``A multiple target Doppler estimation algorithm for OFDM based intelligent radar systems,'' in {\em Proceedings 7th European Radar Conference,} Paris, France, Sept. 2010.

\bibitem{radar_gu}
J. F. Gu, J. Moghaddasi, and K. Wu, ``Delay and Doppler shift estimation for OFDM-based radar-radio (RadCom) system,'' in {\em Proc. IEEE IWS'15,} Shanghai, China, pp. 1-4, March 2015.

\bibitem{radar_tian}
X. Tian, T. Zhang, Q. Zhang, H. Xu, and Z. Song, ``Range and velocity estimation for OFDM-based radar-radio systems,'' in {\em 2017 9th International Conference on Wireless Communications and Signal Processing (WCSP),} Nanjing, pp. 1-6, 2017.

\bibitem{radar_turlapaty}
A. Turlapaty, Y. Jin, and Y. Xu, ``Range and velocity estimation of radar targets by weighted OFDM modulation,'' in {\em Proc. RadarConf'14,} Cincinnati, Ohio, USA, pp. 1358-1362, May 2014.

\bibitem{radar_Wiesbeck}
W. Wiesbeck, ``The radar of the future,'' in {\em Proc. Radar Conference (EuRAD),}
European, pp. 137-140, Oct 2013.

\bibitem{radar_Lellouch}
G. Lellouch, A. Mishra, and M. Inggs, ``Processing alternatives in OFDM
radar,'' in {\em Proc. IET Radar'14,} Lille, France, pp. 1-6, Oct. 2014.

\bibitem{radar_Franken}
G. Franken, H. Nikookar, and P. Van Genderen, ``Doppler tolerance of
OFDM-coded radar signals,'' in {\em Proc. EuRAD'06,} Manchester, UK, pp. 108-111, Sept. 2006.

\bibitem{Hadani}
R.\ Hadani, S.\ Rakib, M.\ Tsatsanis, A.\ Monk, A.\ J.\ Goldsmith, A.\ F.\ Molisch, and R.\ Calderbank, ``Orthogonal time frequency space modulation,'' in {\it Proc. IEEE  Wireless Communications and Networking Conference (WCNC)}, San Francisco, CA, USA, March 2017.

\bibitem{Hadani1}
R. Hadani, S. Rakib, S. Kons, M. Tsatsanis, A. Monk, C. Ibars, J. Delfeld, Y. Hebron, A. J. Goldsmith, A.F. Molisch, and R. Calderbank, ``Orthogonal time frequency space modulation,'' Available online: https://arxiv.org/pdf/1808.00519.pdf.

%\bibitem{otfs_white}
%R. Hadani and A. Monk, ``OTFS: A new generation of modulation addressing the challenges of 5G,'' {\em OTFS %Physics White Paper}, Cohere Technologies, 7 Feb.\ 2018, available online: %https://arxiv.org/ftp/arxiv/papers/1802/1802.02623.pdf.

\bibitem{farhang}
A. Farhang, A. RezazadehReyhani, L. E. Doyle, and B. Farhang-Boroujeny, ``Low complexity modem structure for OFDM-based orthogonal time frequency space modulation,'' in {\em IEEE Wireless Communications Letters,} %vol. 7, no. 3, pp. 344-347, June 2018.

\bibitem{Ravi2}
P. Raviteja, Y. Hong, E. Viterbo, and E. Biglieri, ``Practical pulse-shaping waveforms for reduced-cyclic-prefix OTFS,'' {\em IEEE Trans. on Vehicular Technology}, Oct. 2018, doi: 10.1109/TVT.2018.2878891. 

\bibitem{Ravi1}
P.\ Raviteja, et al., ``Interference cancellation and iterative detection for orthogonal time frequency space modulation,'' {\em IEEE Trans.\ Wireless Commun.}, vol. 17, no. 10, pp. 6501-6515, Oct. 2018.

\bibitem{Ravi}
P.\ Raviteja, et al., ``Low-complexity iterative detection for orthogonal time frequency space modulation,'' in {\em Proc. IEEE Wireless Communications and Networking Conference (WCNC),} Barcelona, April 2018.

\bibitem{Ravi3}
P. Raviteja, K. T. Phan, Y. Hong, and E. Viterbo, ``Embedded delay-Doppler channel estimation for orthogonal time frequency space modulation,'' in {\em Proc. IEEE VTC2018-fall,} Chicago, USA, August 2018.

\bibitem{Ravi4}
P. Raviteja, K. T. Phan, and Y. Hong, ``Embedded pilot-aided channel estimation for OTFS in delay-Doppler channels,'' submitted in {\em IEEE Transactions on Vehicular Technology}.

%\bibitem{Li_Sept2017}
%Li Li, H. Wei, Y. Huang, Y. Yao, W. Ling, G. Chen, P. Li, and Y. Cai, ``A simple %two-stage equalizer With simplified orthogonal time frequency space modulation %over rapidly time-varying channels,'' available online: %https://arxiv.org/abs/1709.02505.

%\bibitem{zemen}
%T. Zemen, M. Hofer, and D. Loeschenbrand, ``Low-complexity equalization for %orthogonal time and frequency signaling (OTFS),'' available online: %https://arxiv.org/pdf/1710.09916.pdf.

%\bibitem{zemen1}
%T. Zemen, M. Hofer, D. Loeschenbrand, and C. Pacher, ``Iterative detection for %orthogonal precoding in doubly selective channels'', available online: %https://arxiv.org/pdf/1710.09912.pdf.

\bibitem{chocks}
K. R. Murali, and A. Chockalingam, ``On OTFS modulation for high-Doppler fading channels,'' in {\em Proc. ITA'2018,} San Diego, Feb. 2018.

\end{thebibliography}
\end{document}